\documentclass[11pt,fleqn]{article}
\usepackage{exscale,amssymb}
\usepackage[dvips]{graphicx}
\usepackage{color}
\usepackage{float}
\usepackage[intlimits]{amsmath}
\usepackage[english]{babel}
\usepackage{rawfonts,latexsym,lscape,epsfig,multirow,lscape,graphics}
\usepackage[justification=raggedright,font=large]{caption}
\usepackage{dcolumn}
\usepackage{rotating}
\usepackage{enumerate}
\usepackage{bbm}

 \newcommand{\bol}[1]{\mbox{\boldmath$#1$}}
\newcommand{\mb}{\mathbf}

\newcommand{\bSigma}{\bol{\Sigma}}

\newcommand{\bm}{\bol{\mu}}

\newcommand{\bx}{\mathbf{x}}
\newcommand{\bQ}{\mathbf{Q}}

\newcommand{\bB}{\mathbf{B}}

\newcommand{\bi}{\mathbf{1}}

\newcommand{\bI}{\mathbf{I}}

\newcommand{\tr}{\mbox{tr}}

\newcommand{\bA}{\bol{A}}

\newcommand{\bF}{\mathbf{\Phi}}

\newcommand{\bxi}{\boldsymbol{\xi}}

\newcommand{\bV}{\mathbf{W}}
\newcommand{\bla}{\boldsymbol{\lambda}}
\newcommand{\bw}{\boldsymbol{\omega}}

\newtheorem{theorem}{Theorem}

\newtheorem{proposition}{Proposition}

\newtheorem{corollary}{Corollary}
\newfont{\tabfont}{cmr7 at 7pt}

\begin{document}
\renewcommand{\baselinestretch}{1.4}
\title{{\bf \vspace*{-1cm} `To Have What They are Having': Portfolio Choice for Mimicking Mean-Variance Savers} \vspace*{2cm}}

\author{
{\bf Vasyl Golosnoy}\thanks{Lehrstuhl f\"{u}r Statistik und \"{O}konometrie, Ruhr-Universit\"{a}t Bochum,
Universit\"{a}tsstra{\ss}e 150, D44780 Bochum, Germany,  phone: +49(0)234-32-22917, fax: +49(0)234-32-14528.
{\sl E-mail:} vasyl.golosnoy@rub.de (the corresponding author)}\\
{\em {\small Lehrstuhl f\"{u}r Statistik und \"{O}konometrie, Ruhr-Universit\"{a}t Bochum, Germany}}\\[0.5cm]
{\bf Nestor Parolya}\thanks{Institut f\"{u}r Statistik, K\"{o}nigsworther Platz 1, 30167 Hannover, phone
 +49(0)511-762-5587, {\sl E-mail:} parolya@statistik.uni-hannover.de}\\
{\em {\small Institut f\"{u}r Statistik, Leibniz Universit\"{a}t Hannover, Germany}}\\[0.5cm]
} 

{\scriptsize \date{(\today)}}

\maketitle 

\begin{abstract}
\noindent
We consider a group of mean-variance investors with mimicking desire such that each investor is willing to penalize deviations of his portfolio composition from compositions of other group members. Penalizing norm constraints are already applied for statistical improvement of Markowitz portfolio procedure in order to cope with estimation risk. We relate these penalties to individuals' wish of social learning and introduce a mutual fund (investment club) aggregating group member preferences unknown for individual savers. We derive the explicit analytical solution for the fund's optimal portfolio weights and show advantages to invest in such a fund for individuals willing to mimic.
\end{abstract}

\normalsize

\vspace{0.7cm}


\newpage

\section{Introduction}

\renewcommand{\baselinestretch}{1.6}


The empirical applicability of the Markowitz mean variance portfolio selection paradigm suffers primarily due to estimation risk concerning model parameters (cf. Klein and Bawa 1976, Best and Grauer 1991). Different methods are suggested in order to repair the Markowitz procedure such as shrinkage estimators (Golosnoy and Okhrin 2009, Frahm and Memmel 2010) or constraining portfolio weights (Jagannathan and Ma, 2003, Behr et al., 2013). Another recent popular approach is to impose an additional norm (quadratic form) constraint on the vector of portfolio weights (Brodie et al., 2009, Fan et al., 2012), which is equivalent to adding a penalty component to the mean-variance objective function. The statistical background of such constraints is related to regularization techniques such as LASSO and/or ridge regression, see Tibshirani (1996, 2011) for recent reviews. From the empirical point of view, imposing constraints on portfolio weights leads to robust portfolio compositions and, consequently, allows to achieve higher expected utility of terminal wealth. Up to now, however, the idea of constraining norm of portfolio weights has not been motivated from economic or decision theoretical perspectives.

In this paper, we suggest that such penalties can be related to a behavioral desire for social learning among the group of investors.
Social learning behavior of an individual means that he wants to act like others (cf. Gilboa et al., 2006), i.e. to mimic portfolio decisions of other group members.  It presumes that each investor is willing to choose portfolio weights by penalizing deviations of his portfolio composition from compositions of others. This could be rational for economic agents in situations where their
own information signals are too noisy. In portfolio selection context it corresponds to the desire `to follow the crowd' where a decision maker wants to hold the same portfolio as the others regardless of his own information signals (cf. Jehiel 2001, Bikhchandani et al., 2008, Park and Sabourian 2011). Mimicking (herding) behavior refers to highly persistent stylized features observed in financial markets (Devenow and Welch 1996, Welch 2000). Recently, Bursztyn et al. (2014), Heimer (2016), Li (2014), Pool et al. (2015) provide broad empirical evidence for various aspects of this phenomenon.
 
In order to formalize the concept of `willing to have what they are having', we modify individual mean-variance objective functions by penalizing deviations from other investors' portfolios, which is defined as a non-negative quadratic norm pre-multiplied with an individual parameter of mimicking desire. This quadratic norm measures the distance between the vector of individual portfolio weights and the vector of (aggregated) portfolio weights of all investors in the group. The individual penalty parameter in our context has a clear economic interpretation as a measure of how unwillingly the saver deviates from the portfolio composition of the rest of the group. Thus, we interpret a desire to `keep up with the Joneses' as an economic motivation for the introduced (penalized) norm constraint for individual mean-variance portfolio investors. The introduced mimicking desire can be seen as a kind of generalization of mean-variance analysis of tracking error (cf. Roll 1992) as we allow for a balance between risk aversion and mimicking (tracking) coefficients.

As in our setting an individual saver is unaware about risk aversions and mimicking coefficients of other group members, the stated portfolio problem remains infeasible for individuals.  In order to overcome this difficulty, we introduce a concept of a mutual fund which functions as an investment club (Barber and Odean 2000) and aggregates preferences of our individual investors (Gilboa et al. 2004, Gollier, 2007). Further, assume that the fund managers are aware via personal communications  about the risk attitude and mimicking desire of individual savers, so that it can exploit this information for resolving the stated optimization task. The role of personal communications for financial decisions is empirically documented e.g. by Hong et al. (2005), Ivkovic and Weisbenner (2007), Li (2014), Pool et al. (2015).
Introducing such fund allows to obtain the compact explicit analytical solution for the penalized portfolio problem.
For this purpose, we first show that the aggregate penalized mean-variance objective function of the mutual fund can be equivalently represented as a mean-variance problem with modified risk attitudes. Then, relying on this representation, we derive the optimal portfolio compositions both for the fund and for individual savers. The resulting vector of the optimal portfolio weights depends on the moments (the expectation vector and the covariance matrix) of asset returns, risk aversions and coefficients of individual mimicking desire. As expected, due to informational advantage of the fund this optimal solution allows to individual investors to achieve higher values of the objective function compared to the case of neglecting their desire to mimic.

The major contributions of the paper can be summarized as follows. First, we relate imposing particular constraints on the norm of portfolio weights to a behavioral phenomenon for social learning which, however, remains infeasible as the investors do not know preferences of each other. Second, we introduce a mutual fund which aggregates individual preferences and derive the analytical expression for the optimal portfolio compositions of individual savers. Finally, we quantify the benefits of pooling money to the mutual fund in case of savers willing to mimic within a comprehensive numerical study.

The rest of the paper is organized as follows. In Section 2, we formulate the mean-variance portfolio problem and introduce a mutual fund for learning and aggregating individual preferences. In Section 3, we formalize mimicking behavior in portfolio selection by imposing an additional norm constraint on the objective function. Then we show that this aggregated penalized objective function can be equivalently written in the form of a mean-variance representation and present the analytical solution of this portfolio problem. In Section 4, we quantify benefits of holding the mutual fund by means of the numerical analysis. Section 5 concludes the paper, whereas the proofs are provided in the Appendix.

\section{Individual investors}

Suppose that there are $k\geq 2$ risky assets with the vector of returns $\bx=(x_1,\ldots,x_k)^\prime$ characterized by the expectation vector $\bm=E(\bx)=(\mu_1,\ldots,\mu_k)^\prime$ and the covariance matrix $\bSigma=Var(\bx)$, which is assumed to be symmetric and positive definite.
Next assume a group of $n\geq 2$ mean-variance investors which possess distinct positive risk aversions $\boldsymbol{\alpha}=(\alpha_1,\ldots,\alpha_n)^\prime$ with $\alpha_i\neq \alpha_j$ for all $i\neq j$ and some initial positive wealth with relative proportions $\boldsymbol{\beta}=(\beta_1,\ldots,\beta_n)^\prime$ such that $\beta_i>0$ and $\sum\limits_{i=1}^n\beta_i=1$.
All individual investors are assumed to have the same information concerning the investment opportunities, i.e. the knowledge of $(\bol{\mu},\mb{\Sigma})$ moments of the asset return distribution. The task is to determine the optimal portfolio composition for these single period individual savers with homogenous beliefs and heterogenous preferences.

\subsection{Individual portfolios}

Denote the vector of portfolio weights of the $i$th investor by $\mb{w}_i$ with $\mb{w}'_i\mb{1}=1$, where $\mb{1}$ is the vector of ones of the corresponding dimension. Accordingly, the portfolio return is given as $r_i=\mb{w}'_i\mb{x}$. Then the individual mean-variance saver maximizes the objective function $EU_i$
\begin{eqnarray}\label{MVi}
\text{max$_{\mb{w}_i}$}~ \left[EU_i=E(r_i)-\dfrac{\alpha_i}{2}Var(r_i)=\mb{w}_i^\prime\bm-\dfrac{\alpha_i}{2}\mb{w}_i^\prime\bSigma\mb{w}_i\right]~ \qquad \text{w.r.t.}  \qquad \mb{w}_i^\prime\bi=1\,.
\end{eqnarray}
 Note that short sellings are allowed in this setting.
The vector of the optimal portfolio weights of the $i$th saver is denoted as $\bw_i=(\omega_i^1,\ldots,\omega_i^k)^\prime$ which is the solution\footnote{We differentiate in notation between a portfolio composition $\mb{w}$ and the optimal portfolio composition
$\bw$.} of the individual optimization task in (\ref{MVi}):
\begin{equation}\label{wMVi}
\bw_i=\bw_{GMV}+\alpha_i^{-1}\bQ\bm\,, \qquad \bQ=\bSigma^{-1}+\dfrac{\bSigma^{-1}\bi\bi^\prime\bSigma^{-1}}{\bi^\prime\bSigma^{-1}\bi},
\end{equation}
where $\bw_{GMV}=\bSigma^{-1}\bi/\bi^\prime\bSigma^{-1}\bi$ is the vector of the global minimum variance portfolio (GMVP) weights. The GMVP is the starting point of the mean-variance efficient frontier with the mean portfolio return
$\mu_{GMV}=\bm'\bSigma^{-1}\bi/\bi'\bSigma^{-1}\bi$ and the variance $V_{GMV}=1/\bi'\bSigma^{-1}\bi$.
The return and variance of the $i$th investor's portfolio are given as $\mu_{i}=\mu_{GMV}+\alpha^{-1}_i \bm' \mb{Q} \bm$ and
$\sigma^2_{i}= V_{GMV}+\alpha^{-2}_i \bm' \mb{Q} \bm$, respectively. This solution corresponds to the mean and variance of a portfolio referring to the efficient frontier. Note that the optimal portfolio proportions $\bol{\omega}_i$ and $\bol{\omega}_j$, for all $i\neq j$, are different only for investors with distinct risk aversions $\alpha_i\neq \alpha_j$ and only if risky assets exhibit distinct expected returns.
If all expected returns are equal such that $\bol{\mu}=\mu\mb{1}$, the GMVP should be selected by all investors independent on their risk aversion coefficients (cf. Bodnar and Okhrin, 2013).

\subsection{Mutual fund}

Now consider a mutual fund which possesses the same information about the available investment opportunities, i.e. about $\bm$ and $\bSigma$. Additionally, the fund knows the risk aversions $\boldsymbol{\alpha}$ of all investors, for example, this information could be obtained via personal communication.
Further assume that the mutual fund just collects money from individual savers so that its own objective function is defined as a wealth-weighted linear combination of individual objective functions (cf. Gilboa et al., 2004):
\begin{eqnarray}\label{eq:noncoopMF}
  EU=\sum_{i=1}^n \beta_i EU_i.
\end{eqnarray}
Note that the only informational advantage of the fund is the knowledge of individual investors' preferences. Thus, our fund can be seen as a kind of investment club (cf. Barber and Odean 2000) where people interchange their opinions and make trading decisions together.

Then the portfolio weights of the fund are given as $\mb{w}_{\mb{f}}=\sum_{i=1}^n \beta_i \mb{w}_i$, or, in the vector notation
$\mb{w}_{\mb{f}}=\mb{W}\bol{\beta}=(\mb{w}_1,...,\mb{w}_n) \bol{\beta}$,
where $k\times n$ matrix $\mb{W}$ summarizes portfolio compositions of all $n$ individual savers.
Accordingly, the fund portfolio return is defined by $r_{\mb{f}}=\bol{\beta}'\mb{W}'\mb{x}$.
Consequently, the optimal (aggregated) portfolio composition of the mutual fund is given as the solution of the task $\max_{\mb{w}_{\mb{f}}} EU$ w.r.t. $\mb{w}'_{\mb{f}}\mb{1}=1$:
\begin{eqnarray}\label{noncoop}
\bw_{\mb{f}}=\bw_{GMV}   + \alpha^{-1}_{\mb{f}} \bQ \bm, \qquad \alpha_{\mb{f}} = (\mb{1}' \mb{A}_0^{-1} \bol{\beta})^{-1},
\qquad \mb{A}_0=\mbox{diag}(\alpha_1,...,\alpha_n),
\end{eqnarray}
where $\alpha_{\mb{f}}$ is the (aggregated) risk aversion of the fund.
The mean and the variance of the fund portfolio return are given, respectively, as
$\mu_{\mb{f}}= \mu_{GMV} + \alpha^{-1}_{\mb{f}} \bm'\mb{Q}\bm$ and
$\sigma^2_{\mb{f}}= V_{GMV}+ \alpha^{-2}_{\mb{f}} \bm'\mb{Q}\bm$. The optimal portfolio compositions of individual investors are contained in the matrix $\bol{\mathcal{W}}$ with $\bw_{\mb{f}}=\bol{\mathcal{W}}\bol{\beta}=(\bw_1,...,\bw_n) \bol{\beta}$.

We assume that the information about risk aversions of individual savers is available to the mutual fund but not to individual investors which are not aware about  preferences of each other. It is easy to see, however, that the mutual fund cannot bring additional advantages for individual savers in this setting, because its optimal portfolio compositions is merely a weighted average (linear combination) of the individual optimal portfolio holdings. This result, however, would be different under assumption that individual savers incline to mimic behavior of each other. We consider this case in the next section.


\section{Investors with a desire to mimic portfolio compositions}

The mutual fund introduced in previous section has an information advantage compared to individual savers due to the knowledge of risk aversions.
In this section, we show that this informational advantage appears to be relevant for investment decisions in cases where individuals do not want to hold portfolios which strongly deviate from aggregated holdings of others.
Since this behavior implies readiness to account for decisions of others we denote this phenomenon as a desire to mimic in a portfolio selection
context. The behavioral explanation of this phenomenon roots in the well-documented human inclination to mimic actions (decisions) of others (cf. Devenow and Welch, 1996) as well as on social learning arguments (Park and Sabourian 2011, Bursztyn et al., 2014). In our setting investors are unaware about preferences of each other, so that they cannot mimic themselves but availability of the mutual fund allows to resolve this problem.

\subsection{Mimicking and penalized utility function}

Now, we introduce the mean variance portfolio problem with mimicking by adjusting individual objective functions such that they incorporate a penalty component. In particular, for the $i$th investor we penalize deviations of the
investor's portfolio composition from the aggregate composition of the mutual fund which contains all individual portfolios in the group.
Then the optimization task of the $i$th investor has the following representation
\begin{eqnarray}\label{cMVi}
 \max_{\mb{w}_i} \, \left[EU^*_i=EU_i-\dfrac{\phi_i}{2}(\mb{w}_i-\mb{w}_{\mb{f}})^\prime\bSigma(\mb{w}_i-\mb{w}_{\mb{f}})\right], \quad \mb{w}_{\mb{f}}=\mb{W}\bol{\beta}, \quad \mbox{w.r.t.} \quad \mb{w}'_i\mb{1}=1,
\end{eqnarray}
where the penalty factor $\phi_i>0$ is an individual coefficient of mimicking inclination. A large value of $\phi_i$ implies that the $i$th investor is definitely willing to hold a portfolio composition which is `close' to compositions of others, i.e. exhibits a stronger desire to mimic. This is in line with Ahern et al. (2014) who find out that peer effects may lead to changes in risk aversions of group participants.
Using the covariance matrix $\bSigma$ for weighting the distance allows us to remain in the mean-variance framework which simplifies substantially  the solution of this portfolio problem. Note that the case without mimicking desire $\phi_i=0$ for $i=1,...,n$ corresponds to the classical mean-variance framework which is considered in Section 2.

It is known that penalty components as introduced in Equation (\ref{cMVi}) are useful for a practical portfolio choice.
Imposing various constraints on portfolio weights are recently applied in practical portfolio selection by (among others) Jagannathan and Ma (2003), Brodie et al., (2009). These constrains were suggested in order to cope with estimation risk in the portfolio weights by relying on ideas behind robust statistical techniques, such as LASSO or ridge regression (cf. Tibshirani et al., 1996, 2011). Thus, up to now the usefulness of constraints for portfolio weights is motivated solely by statistical argumentation.

In this paper, we reveal the economic motivation for imposing constraints as in Equation (\ref{cMVi}). In particular, we claim that the intuition behind introduction of such penalty function is twofold. Firstly, penalizing deviations in portfolio composition is related to a social learning phenomenon in financial markets where agents are eager to change their beliefs and behavior as a result of observing the actions of others (Park and Sabourian, 2011). Empirically, there is a well-documented behavioral desire to mimic in financial market (cf. Devenow and Welch, 1996) which relies on similarity arguments and is rather fundamental for human beings (cf. Gilboa et al., 2006). Secondly, a desire not to be overperformed by peers (e.g., `keeping up with the Joneses' argument) would also impose mimicking desire of an individual saver by portfolio composition. Thus, imposing constraints on portfolio weights introduced in the mean-variance optimization task (\ref{cMVi}) has not only statistical but also economical motivation.


\subsection{Mean-variance representation for investors with mimicking desire}

The mimicking desire is formalized above by penalizing the distance between an individual portfolio composition and the aggregated holdings of the mutual fund. Since the investor $i$ remains unaware about risk aversions and mimicking coefficients of other investors in the group, the optimal solution of the problem in (\ref{cMVi}) appears to be individually infeasible. Since the mutual fund possesses this information, it can help individual investors to find the optimal solution of (\ref{cMVi}). Next we show analytically that individual investors with mimicking desire would get advantage by having such fund (investment club) at hand.

As earlier, we aggregate objective functions of individuals willingness to mimic, so that the objective function is given as:
\begin{eqnarray}\label{eq:eu*}
   \max_{\mb{w}_{\mb{f}}} \left[ EU^* = \sum_{i=1}^n \beta_i EU^*_i\right], \qquad \mbox{w.r.t.} \qquad \mb{w}'_{\mb{f}}\mb{1}=\bol{\beta}'\mb{W}'\mb{1}=1.
\end{eqnarray}
The idea to aggregate portfolio holdings allows to obtain the explicit solution for the optimal portfolio, which cannot be achieved without this step (cf. Brodie et al., 2009).

In order to solve analytically the optimization task in (\ref{eq:eu*}), we rewrite the aggregate optimization problem such that it has a mean-variance representation. Denote the vector of individual mimicking coefficients by $\boldsymbol{\phi}=(\phi_1,\ldots,\phi_n)^\prime$ and the corresponding diagonal matrix as $\boldsymbol{\Phi}=\text{diag}\left\{\phi_1,\ldots,\phi_n\right\}$. The diagonal matrices $\mb{A}_0=\text{diag}\left\{\alpha_1,\ldots,\alpha_n\right\}$ and $\bol{B}=\text{diag}\left\{\beta_1,\ldots,\beta_n\right\}$ contain information about the risk aversion coefficients and the relative wealth of each investor, respectively. The mean-variance form of the  mutual fund objective function given in (\ref{eq:eu*}) is provided in the following proposition.

\begin{proposition} The aggregated objective function of \texttt{n} mimicking investors $EU^*$ defined in (\ref{cMVi}) and (\ref{eq:eu*}) has the following equivalent mean-variance representation:
\begin{eqnarray}\label{equivEU}
EU^*&=&E(\bol{\beta}'\mb{W}'\mb{x})-\dfrac{1}{2}E\left[\left(\mb{x}-E(\mb{x})\right)^\prime \mb{W} \bA \mb{W}'\left(\mb{x}-E(\mb{x})\right)\right]\nonumber\\
&=&\boldsymbol{\beta}^\prime\bV^\prime\bm-\dfrac{1}{2}\tr(\bA\bV^\prime\bSigma\bV)\,,
\end{eqnarray}
where the mimicking matrix $\bA$ of dimension $n\times n$ is given by
\begin{equation}
\bA=\left(\mb{A}_0+\bF\right)\bol{B}+\left(\bar{\phi}\bI-2\bF\right)\bol{\beta}\bol{\beta}'\,, \qquad \bar{\phi}=\boldsymbol{\beta}^\prime\boldsymbol{\phi}=\sum\limits_{i=1}^n\beta_i\phi_i.
\end{equation}
\end{proposition}

The proof of Proposition 1 is provided in the Appendix. The statement of Proposition 1 allows to obtain the solution of the mutual fund mean-variance problem by using the standard portfolio optimization toolkit, as it is done below.

\subsection{Solution of portfolio problem with mimicking desire}

Since the return on the mutual fund portfolio is given as $r_{\mb{f}}=\mb{w}'_{\mb{f}}\mb{x} = \bol{\beta}' \mb{W}' \mb{x}$,
our task is to find the optimal portfolio weight $k\times n$ matrix $\bol{\mathcal{W}}^*$ which contains the information about all individual portfolios. The corresponding optimization problem
\begin{eqnarray}\label{mv1}
\max_{\bV} \left[EU^*=\boldsymbol{\beta}^\prime\bV^\prime\bm-\dfrac{1}{2}\tr(\bA\bV^\prime\bSigma\bV)\right] \qquad \mbox{w.r.t.}\qquad
\bV^\prime\bi_k=\bi_n,
\end{eqnarray}
is solved analytically in Theorem 1 which is proven in the Appendix.

\begin{theorem}
 Consider \texttt{n} individual investors with the aggregate objective functions given in (\ref{mv1}). Then the optimal mimicking portfolio weights are given by
\begin{equation}\label{MVweights}
\bol{\mathcal{W}}^*=\dfrac{\bSigma^{-1}\bi}{\bi^\prime\bSigma^{-1}\bi}\bi_n^\prime+
\bQ\bm(\boldsymbol{\beta}^\prime \bA_{\phi}^{-1})\,, \qquad \bA_{\phi}=(\bA+\bA')/2,
\end{equation}
where the matrix $\bA_{\phi}$ is symmetric and positive definite.
\end{theorem}

Since it holds that $\bol{\mathcal{W}}^*=(\bol{w}^*_1,...,\bol{w}^*_n)$ and $\bol{w}^*_{\mb{f}}=\bol{\mathcal{W}}^* \bol{\beta}$
the mimicking (aggregated) optimal weights of the fund are given as
\begin{eqnarray}\label{eq:coopMF}
  \bw^*_{\mb{f}}=\dfrac{\bSigma^{-1}\bi}{\bi^\prime\bSigma^{-1}\bi}+(\alpha^*_{\mb{f}})^{-1} \bQ\bm , \qquad
  (\alpha^*_{\mb{f}})^{-1}=\boldsymbol{\beta}^\prime\bA^{-1}_{\phi}\boldsymbol{\beta},
\end{eqnarray}
whereas the mean and variance of the fund portfolio return in the case of mimicking are
$\mu^*_{\mb{f}}=\mu_{GMV}+\left(\boldsymbol{\beta}^\prime\bA^{-1}_{\phi}\boldsymbol{\beta}\right)\bm^\prime\bQ\bm$ and
$\sigma^{2\,*}_{\mb{f}}=V_{GMV}+\left(\boldsymbol{\beta}^\prime\bA^{-1}_{\phi}\boldsymbol{\beta}\right)^2 \bm^\prime\bQ\bm$, respectively.
Thus, the mimicking behavior has impact only on the aggregate risk aversion coefficient $\alpha^*_{\mb{f}}$.

The most important special cases of Theorem 1 are summarized in Corollary 1. In particular, it provides the solutions of the portfolio problem for the cases of (a) equal wealth for all investors $\beta_i=1/n$; (b) equal mimicking coefficients for all investors $\phi_i=\phi>0$; (c) equal risk aversions and equal mimicking coefficients for all investors.


\begin{corollary} The important special cases resulting from Theorem 1 are given as:
\begin{enumerate}[a)]
\item Suppose equal wealth of all investors $\beta_i=1/n$.
 Then the mimicking matrix $\bA$ is given by
\begin{equation}
\bA=n\left(\mb{A}_0+\bF\right)+\left(\bar{\phi}\bI-2\bF\right)\bi_n\bi_n^\prime\,
\end{equation}
and the corresponding optimal weights of the fund portfolio are equal to
\begin{equation}
\bw^*_{\mb{f}}=\dfrac{\bSigma^{-1}\bi}{\bi^\prime\bSigma^{-1}\bi}+\bQ\bm \; \left(\bi_n^\prime\bA^{-1}_{\phi}\bi_n\right) .
\end{equation}%

\item Suppose equal mimicking coefficients for all investors, i.e. $\phi_i=\phi>0$ for all $i\in\{1,\ldots,n\}$. Then the symmetric mimicking matrix $\bA$ is always positive definite and given by
\begin{equation}
\bA=\left(\mb{A}_0+\phi\bI\right)\bB-\phi\bol{\beta}\bol{\beta}'\,.
\end{equation}
The corresponding optimal portfolio weights of the fund are given as
\begin{equation}
\bw^*_{\mb{f}}=\dfrac{\bSigma^{-1}\bi}{\bi^\prime\bSigma^{-1}\bi}+\bQ\bm \; \left(\bol{\beta}^\prime\bA^{-1}\bol{\beta}\right).
\end{equation}
\item Assume that all investors have the same risk aversion and mimicking coefficients, i.e. $\alpha_i=\alpha$ and $\phi_i=\phi$ for all $i\in\{1,\ldots,n\}$. Then the optimal portfolio weights for savers willing to mimic coincide with a non-mimicking mean-variance portfolio composition given by (\ref{noncoop}).
\end{enumerate}
\end{corollary}

The results of Corollary 1 imply that the desire to mimic is primarily reasonable for comparative small values of risk aversions and mimicking coefficients as well as for the positive norm $\bol{\mu}'\mb{Q}\bol{\mu}$. Otherwise the optimal portfolio composition would be the GMVP, which is the same for all investors.

As in practice the true distribution parameters $\bol{\mu},\mb{\Sigma}$ are unknown and should be estimated, the estimation risk need to be taken into account in a portfolio selection procedure. Since the (deterministic) mimicking desire enters only the expression for the aggregated risk aversion of the mutual fund $\alpha_{\mb{f}}$, all statistical results derived by Okhrin and Schmid (2006), Bodnar and Schmid (2011) apply directly for quantifying estimation risk in the optimal portfolio weights.

\subsection{Many small investors}

A particular interesting case is to consider $n\to\infty$ small investors. Assuming that $\sup\limits_i \beta_i\rightarrow0$ for $n\rightarrow\infty$
which is very natural as it covers $\beta_i=1/n$ as a special case, we immediately obtain for
$\bA_{\phi}=(\bA+\bA^\prime)/2$ with $\bA=(\mb{A}_0+\bF)\bol{B}+(\bar{\phi}\bI-2\bF)\bol{\beta}\bol{\beta}'$ that
\begin{eqnarray*}
  \lim_{n\to\infty} \bA_{\phi}=(\mb{A}_0+\bF)\bol{B},
\end{eqnarray*}
because the part with $\bol{\beta}\bol{\beta}'$ appears to be of smaller order. This allows to get the asymptotic formula for the aggregate risk  aversion coefficient:
\begin{equation}\label{approx1}
 \lim_{n\to\infty} \left[\alpha^{-1\; *}_{\mb{f}}=\bol{\beta}' \bA_{\phi}^{-1}\bol{\beta}\right]= \sum_{i=1}^n \dfrac{\beta_i}{\alpha_i+\phi_i}.
\end{equation}
This result has a couple of interesting implications. Using this approximation and Jensen's inequality we are able now to find the asymptotic upper bound for the aggregate risk aversion coefficient $\alpha^*_{\mb{f}}$, namely
\begin{equation}\label{ineq11}
 \lim_{n\to\infty} \alpha^*_{\mb{f}} = \left(\sum_{i=1}^n \dfrac{\beta_i}{\alpha_i+\phi_i}\right)^{-1} \leq \sum_{i=1}^n\beta_i\alpha_i+\sum_{i=1}^n\beta_i\phi_i = \bar{\alpha}+\bar{\phi}\,.
 \end{equation}
Thus, investing into the mutual fund is reasonable for any $\bar{\alpha}+\bar{\phi}$ sufficiently small otherwise the GMV portfolio is of interest.

Moreover, combining \eqref{approx1} and inequality \eqref{ineq11} gives us the upper and lower bounds for the aggregated risk aversions of mimicking investors which are given by
\begin{equation}
 \bar{\alpha} + \bar{\phi} > \alpha^*_{\mb{f}}  >  \alpha_{\mb{f}}~~\text{  for sufficiently large $n$}\,.
\end{equation}
This implies that for a large number of investors $n$ with a social learning desire the aggregate risk aversion of a mutual fund increases compared to the non-mimicking case. This fact seems to be natural from the economic point of view as the mimicking desire takes into account the anxiety of the investors about the future uncertain outcomes. As a result, this anxiety drives the investors' wish to be closer to the portfolio decisions of the `rest of the group'.

\section{Numerical illustration}

Using the results of Theorem 1 and Corollary 1, we illustrate benefits of investing into the mutual fund for mean-variance portfolio investors willing to hold similar compositions. For this purpose we conduct a numerical study where we contrast portfolio investments with and without accounting for mimicking desire. In particular, we quantify changes in the optimal portfolio proportions as well as gains in the investor's objective function for different values of risk aversion and mimicking coefficients.

The study is designed as follows. We concentrate on a parsimonious case with $k=2$ risky assets and $n=2$ investors which are characterized by distinct positive risk aversion coefficients $\alpha_1\neq\alpha_2$. Both investors are assumed to have the equal amount of money, i.e. $\beta=1/2$. The distributional parameters of asset returns are selected such that the annualized means of risky assets are equal to $\mu_1=0.07$, $\mu_2=0.14$, the standard deviations $\sigma_1=0.12$ and $\sigma_2=0.2$, the correlation $\rho=0.2$. This choice of distribution parameters is conform to the typical values used in finance textbooks for illustration purposes (cf. Copeland et al., 2004).

Our aim is to investigate the impact of ($\phi_1,\phi_2$) and ($\alpha_1,\alpha_2$) on the optimal portfolio composition in case of individual savers willing to mimic. We concentrate on analyzing the case of equal mimicking coefficients $\phi_1=\phi_2=\phi>0$ whereas the case of $\phi_1\neq \phi_2$ could be investigated in a rather similar line.
In the first step, we select the mimicking coefficient $\phi$ as a value from the set $\{3,5,10\}$ for the given risk aversion $\alpha_1=2$. Then we vary the relation between risk aversions $a=\alpha_2/\alpha_1$ such that $a\in[1,10]$. These risk aversions are typical values applied in the mean-variance portfolio analysis (cf. Golosnoy and Okhrin 2009). In the second step, we consider portfolio choices for the given values of the risk aversion relation $a\in\{2,5,10\}$ and vary the mimicking coefficient $\phi$ such that $\phi\in[0,5]$. Thus we analyze the cases where the mimicking and risk aversion coefficients are of the same order.

\begin{figure}[p!]
\vspace{-5mm}
\includegraphics[scale=0.5]{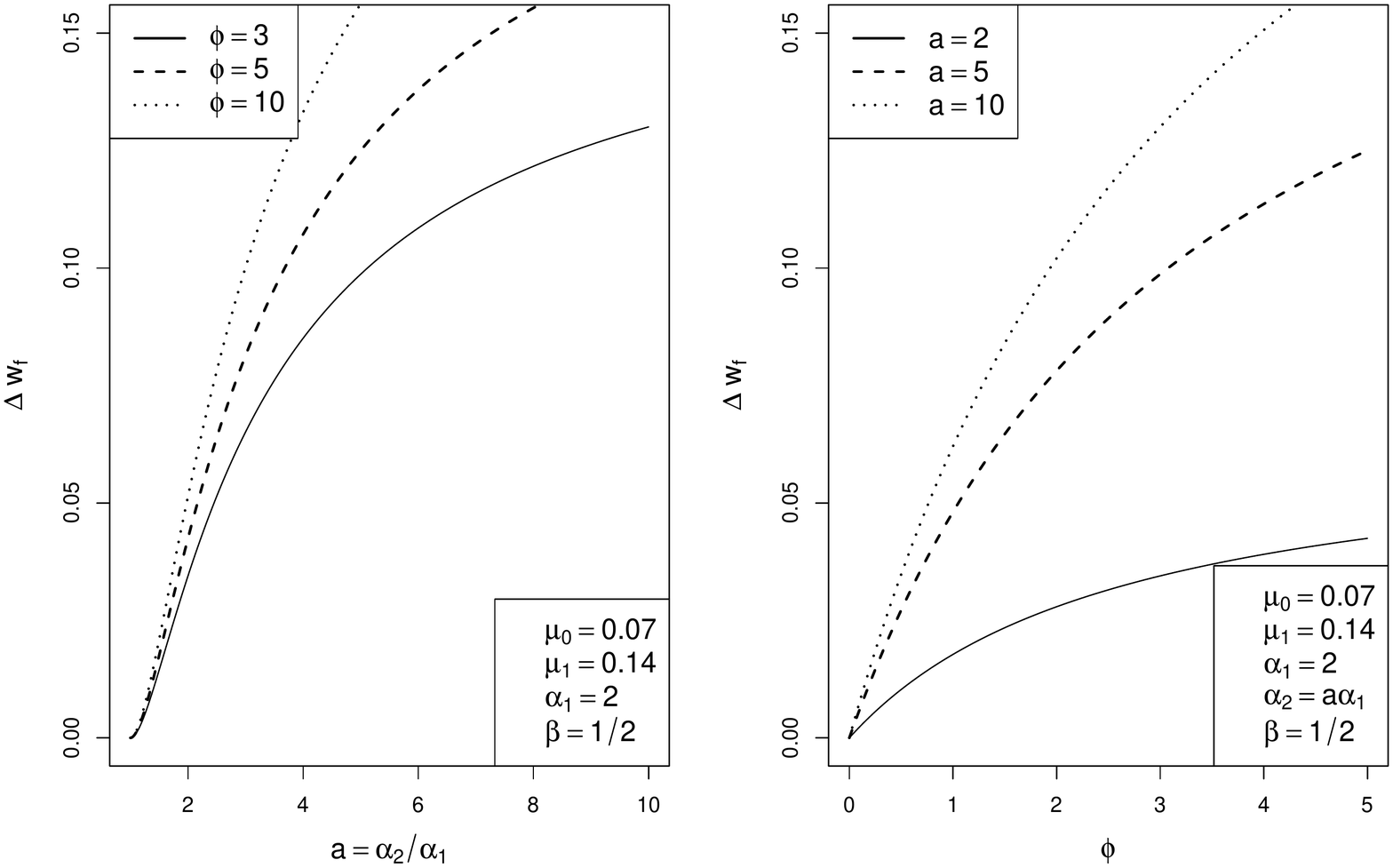}
\caption{{\small Change in the first optimal weight $\Delta \omega$ as a function of $a=\alpha_2/\alpha_1$ (left) and $\phi$ (right) for two-asset portfolio.}}
\label{Fig:1}
\end{figure}

The mimicking behavior should have impact on both optimal portfolio weights and the objective function.
Alterations in the optimal portfolio composition are quantified by considering the difference in the optimal weight of the first asset with mimicking (feasible for the mutual fund) and without mimicking (feasible for individual savers):
\begin{eqnarray*}
\Delta \omega=\omega^*_{\mb{f}}-\omega_{\mb{f}}.
\end{eqnarray*}
The optimal fund weights are given in Equation (\ref{eq:coopMF}) whereas no-mimicking weights in Equation (\ref{noncoop}).
The changes $\Delta \omega$ are plotted in Figure \ref{Fig:1} as functions of $(\phi,a)$.

Denote by $EU^*(\bol{w}^*_{\mb{f}})$ the maximal value of the objective function with mimicking desire given in Equation (\ref{eq:eu*}) which can be attained by the mutual fund. Next, substitute into (\ref{eq:eu*}) the sub-optimal solution $\bol{w}_{\mb{f}}$ and obtain the value $EU^*(\bol{w}_{\mb{f}})$ which is attainable for
individual investors. The corresponding solutions $\bol{w}^*_{\mb{f}}$ and $\bol{w}_{\mb{f}}$ are provided in Equations (\ref{eq:coopMF}) and (\ref{noncoop}), respectively.
The relative gains in the objective function for investors with mimicking desire due to investing into the mutual fund are defined as
\begin{eqnarray}\label{eq:gainsOF}
\Delta EU &=& \frac{EU^*(\bol{w}^*_{\mb{f}})- EU^*(\bol{w}_{\mb{f}})}{EU^*(\bol{w}^*_{\mb{f}})}.
\end{eqnarray}

\begin{figure}[p!]
\vspace{-5mm}
\includegraphics[scale=0.5]{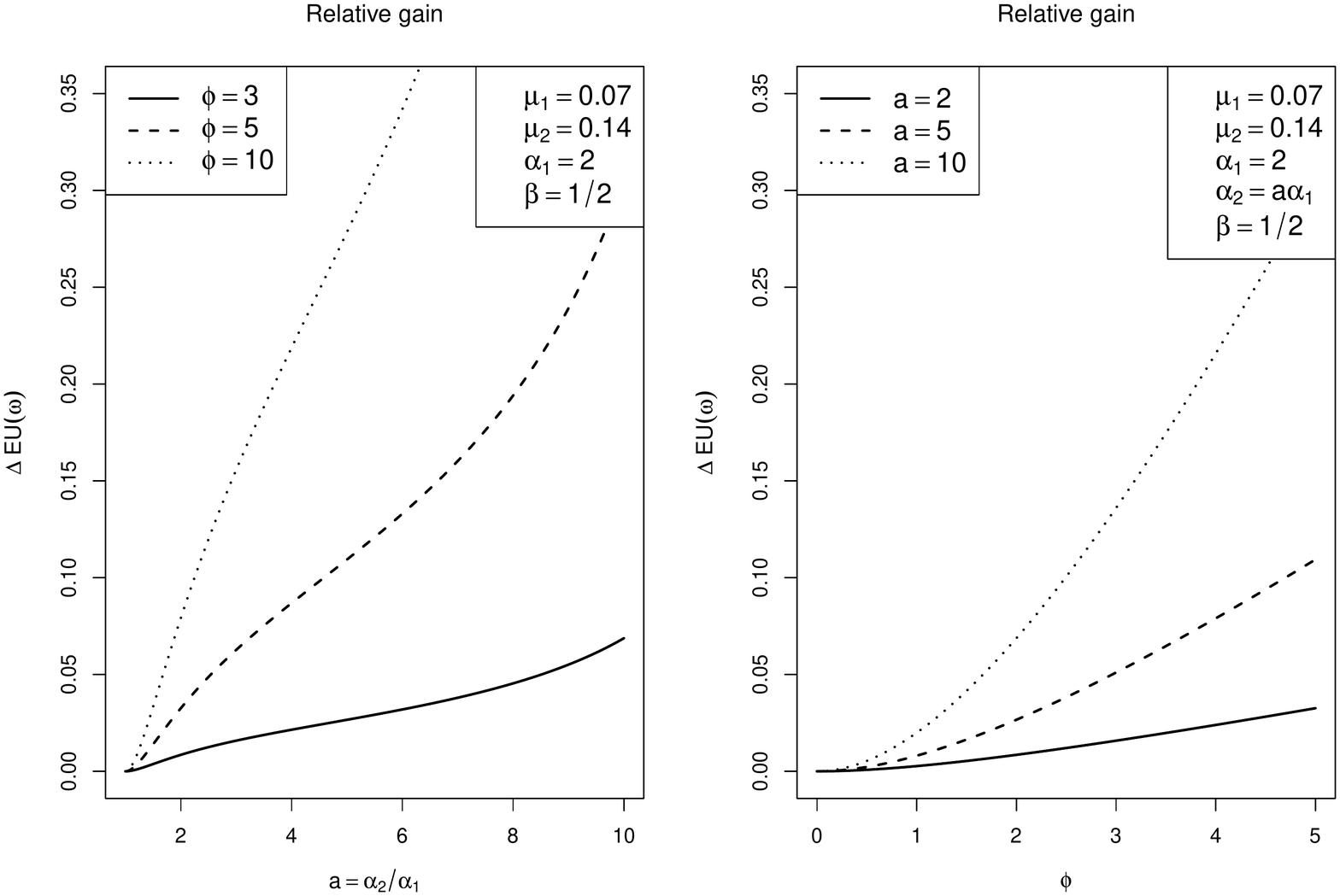}
\caption{{\small Relative utility gain $\Delta EU$ as function of $a=\alpha_2/\alpha_1$ (left) and $\phi$ (right) for two-asset portfolio.}}
\label{Fig:2}
\end{figure}

The relative utility gains $\Delta EU$, plotted in Figure \ref{Fig:2} as a function of $(\phi,a)$ correspond to benefits from mimicking which can be achieved by investing into the mutual fund.

\begin{center}
  [Figures 1 and 2 about here.]
\end{center}

In Figure 1 we observe a non-linear dependence between change in the optimal composition $\Delta \omega$ and $(\phi,a)$.
The change in the optimal portfolio composition appears to be remarkable (over 10\%) already for $\phi\geq 3$ and $a\geq 5$, which are quite realistic values of the mimicking and risk aversion coefficients.

The gains in the objective function (\ref{eq:gainsOF}), visualized in Figure 2, are monotone increasing in $(\phi,a)$. They are clearly pronounced already for quite small values of the mimicking coefficient $\phi$, for example, the gains exceed the level of 10\% for $a\geq 5$, $\phi\geq 5$.
These results show that pooling money into the mutual fund leads to quite remarkable benefits for individual investors with a willingness to mimic and justifies the fees which the fund can take for these advantages. Thus, we establish another reason to invest into a mutual fund additionally to the well-recognized advantages such as diversification and liquidity benefits and/or professional performance by searching and processing relevant information (see e.g. Stracca 2006, Huang et al., 2011).

\section{Summary}

In this paper, we consider mean-variance investors with distinct risk aversion coefficients and mimicking desire which corresponds to a wish to hold portfolios not very different from portfolio compositions of others.
We show that such mimicking inclination corresponds to imposing additional constraints on the norm of portfolio weights, which is used in the recent literature in order to improve performance of the classical Markowitz mean-variance procedure.
Economically,  portfolio behavior with willingness to mimic corresponds to the well-known behavioral desire of
social learning in the sense of staying close to portfolio decisions of others.

Since individual investors are usually unaware about preferences of each others, the portfolio task with mimicking desire remains infeasible to an individual saver. For this reason we introduce a mutual fund which aggregates objective functions of individual investors and possesses information about their risk attitudes and mimicking inclination coefficients. Using these informational advantages, we derive the explicit solution for the vector of the optimal mimicking portfolio weights both for the fund and for individual savers. Thus, we show that pooling money to the mutual fund provides additional benefits for investors willing to mimic investment decisions of their peers.


\section*{Appendix}

\noindent {\bf Proof of Proposition 1.}
It holds that
\begin{eqnarray}\label{refundEU}
EU&=&\sum\limits_{i=1}^n\beta_i EU^*_i=\sum\limits_{i=1}^n\beta_i\left(\bw_i^\prime\bm-\dfrac{\alpha_i}{2}\bw_i^\prime\bSigma\bw_i-\dfrac{\phi_i}{2}(\bw_i-\bw_{\mb{f}})^\prime\bSigma(\bw_i-\bw_{\mb{f}})\right)\\
&=&\sum\limits_{i=1}^n\beta_i\bw_i^\prime\bm-\sum\limits_{i=1}^n\beta_i\dfrac{\alpha_i}{2}\bw_i^\prime\bSigma\bw_i-\sum\limits_{i=1}^n\beta_i\dfrac{\phi_i}{2}\left(\bw_i-\sum\limits_{k=1}^n\beta_k\bw_k\right)^\prime\bSigma\left(\bw_i-\sum\limits_{k=1}^n\beta_k\bw_k\right)\nonumber\,.
\end{eqnarray}
Denote $\bar{\phi}=\sum\limits_{i=1}^n\beta_i\phi_i$ and rewrite the last term in (\ref{refundEU}) as
\begin{eqnarray}\label{MDist}
&&\sum\limits_{i=1}^n\beta_i\phi_i\left(\bw_i-\sum\limits_{k=1}^n\beta_k\bw_k\right)^\prime\bSigma\left(\bw_i-\sum\limits_{k=1}^n\beta_k\bw_k\right)\nonumber\\
&=&\sum\limits_{i=1}^n\beta_i\phi_i\bw_i^\prime\bSigma\bw_i+\bar{\phi}\sum\limits_{i=1}^n\beta^2_i\bw_i^\prime\bSigma\bw_i+\bar{\phi}\underset{i\neq j}{\sum\limits_{{i=1}}^n\beta_i\sum\limits_{j=1}^n}\beta_j\bw_i^\prime\bSigma\bw_j-2\sum\limits_{i=1}^n\beta^2_i\phi_i\bw_i^\prime\bSigma\bw_i-2\underset{i\neq j}{\sum\limits_{i=1}^n\beta_i\sum\limits_{j=1}^n}\beta_j\phi_i\bw_i^\prime\bSigma\bw_j\nonumber\\
&=&\sum\limits_{i=1}^n\beta^2_i\left[\left((\beta^{-1}_i-2)\phi_i+\bar{\phi}\right)\right]\bw_i^\prime\bSigma\bw_i+\underset{i\neq j}{\sum\limits_{i=1}^n\sum\limits_{j=1}^n}\beta_i\beta_j(\bar{\phi}-2\phi_i)\bw_i^\prime\bSigma\bw_j\,,
\end{eqnarray}
Combining (\ref{refundEU}) and (\ref{MDist}) we receive
\begin{eqnarray}\label{rerefundEU}
EU=\sum\limits_{i=1}^n\beta_i\bw_i^\prime\bm-\dfrac{1}{2}\sum\limits_{i=1}^n\beta^2_i\left(\beta_i^{-1}\alpha_i+(\beta_i^{-1}-2)\phi_i+\bar{\phi}\right)\bw_i^\prime\bSigma\bw_i
-\dfrac{1}{2}\underset{i\neq j}{\sum\limits_{i=1}^n\sum\limits_{j=1}^n}\beta_i\beta_j\left(\bar{\phi}-2\phi_i\right)\bw_i^\prime\bSigma\bw_j\,.
\end{eqnarray}

Taking into account that
$Cov(\bw_i^\prime\bx, \bw_j^\prime\bx)=\bw_i^\prime Cov(\bx, \bx)\bw_j=\bw_i^\prime\bSigma\bw_j\,$,
we rewrite (\ref{rerefundEU}) and get
\begin{eqnarray}\label{rerefundEU2}
EU&=&\sum\limits_{i=1}^n\beta_iE\left[\bw_i^\prime\bx\right]-\dfrac{1}{2}\sum\limits_{i=1}^n a_iVar(\bw_i^\prime\bx)-\dfrac{1}{2}\underset{i\neq j}{\sum\limits_{i=1}^n\sum\limits_{j=1}^n} a_{ij}Cov(\bw_i^\prime\bx, \bw_j^\prime\bx)\nonumber\\
&=&\sum\limits_{i=1}^n\beta_iE\left[\bw_i^\prime\bx\right]-\dfrac{1}{2}\sum\limits_{i=1}^n \sum\limits_{j=1}^n a_{ij}Cov(\bw_i^\prime\bx)
=E(\boldsymbol{\beta}^\prime\bV^\prime\bx)-\dfrac{1}{2}E\left[\left(\bx-\bm\right)^\prime\bV\bA\bV^\prime\left(\bx-\bm\right)\right]
\end{eqnarray}
where $a_i=\beta^2_i(\beta_i^{-1}\alpha_i+(\beta_i^{-1}-2)\phi_i+\bar{\phi})$ and $a_{ij}=\beta_i \beta_j (\bar{\phi}-2\phi_i)$ are the diagonal and off-diagonal elements of the matrix $\bA=\left(\mb{A}_0+\bF\right)\bB+\left(\bar{\phi}\bI-2\bF\right)\bol{\beta}\bol{\beta}'$, respectively. Noting that
$E(\boldsymbol{\beta}^\prime\bV^\prime\bx)=\boldsymbol{\beta}^\prime\bV^\prime\bm$ and
\begin{eqnarray}
E\left[\left(\bx-\bm\right)^\prime\bV\bA\bV^\prime\left(\bx-\bm\right)\right]
=\text{tr}\left(\bV\bA\bV^\prime E\left[\left(\bx-\bm\right)\left(\bx-\bm\right)^\prime\right]\right)
=\text{tr}\left(\bV\bA\bV^\prime\bSigma\right)\, ,
\end{eqnarray}
we get the statement of Proposition 1.

\noindent{\bf Proof of Theorem 1.}
Rewriting (\ref{mv1}) via Kronecker product and  vec-operator (Harville, 1997) we receive
\begin{eqnarray}\label{vecMV}
&&vec(\bV^\prime)^\prime(\bm\otimes\bI_n)\boldsymbol{\beta}-\dfrac{1}{2}vec(\bV^\prime)^\prime(\bSigma\otimes\bA)vec(\bV^\prime)\rightarrow\text{max}~~\text{over}~~vec(\bV^\prime)\\
&&\text{subject to} \qquad (\bi^\prime\otimes\bI_n)vec(\bV^\prime)=\bi_n\nonumber\,,
\end{eqnarray}

The first order condition for the optimization problem (\ref{vecMV}) is given by
\begin{equation}\label{Lag2}
(\bm\otimes\bI_n)\boldsymbol{\beta}-\dfrac{1}{2}(\bSigma\otimes(\bA+\bA^\prime))vec(\bV^\prime)-(\bi\otimes\bI_n)\bla=0\,,
\end{equation}
where $\bla$ is $n$-dimensional vector of Lagrange multipliers. Denote $\bA_\phi=1/2(\bA+\bA^\prime)$ then the vector of optimal weights equals
\begin{eqnarray}\label{wmv}
vec(\bV^\prime)&=&(\bSigma\otimes\bA_\phi)^{-1}(\bm\otimes\bI_n)\boldsymbol{\beta}-(\bSigma\otimes\bA_\phi)^{-1}(\bi\otimes\bI_n)\bla\nonumber\\
&=&(\bSigma^{-1}\bm\otimes\bA_\phi^{-1})\boldsymbol{\beta}-(\bSigma^{-1}\bi\otimes\bA_\phi^{-1})\bla\,.
\end{eqnarray}
The vector $\bla$ can be rewritten in terms of Kronecker product and vec-operator in the following way $\bla=vec(\bla)=vec(\bI_n\bla)=(\bla^\prime\otimes\bI_n)vec(\bI_n)$. We do the same with $\boldsymbol{\beta}$ and get
\begin{eqnarray}\label{w1}
vec(\bV^\prime)&=&(\bSigma^{-1}\bm\otimes\bA_\phi^{-1})(\boldsymbol{\beta}^\prime\otimes\bI_n)vec(\bI_n)-(\bSigma^{-1}\bi\otimes\bA_\phi^{-1})(\bla^\prime\otimes\bI_n)vec(\bI_n)\nonumber\\
&=&\left[\left(\bSigma^{-1}\bm\boldsymbol{\beta}^\prime+\bSigma^{-1}\bi\bla^\prime\right)\otimes\bA_\phi^{-1}\right] vec(\bI_n)\,.
\end{eqnarray}
Consequently it holds that
$vec(\bV^\prime)=vec\left[\bA_\phi^{-1}\left(\bSigma^{-1}\bm\boldsymbol{\beta}^\prime+\bSigma^{-1}\bi\bla^\prime\right)^\prime\right]\,$, so  we receive
\begin{equation}\label{mvw3}
\bV^\prime=\bA_\phi^{-1}\left(\boldsymbol{\beta}\bm^\prime\bSigma^{-1}+\bla\bi^\prime\bSigma^{-1}\right)\,.
\end{equation}

After post-multiplying both sides of (\ref{mvw3}) with $\bi$, we solve the task with respect to $\bla$ and get
\begin{equation}\label{lam}
\bla=\dfrac{1}{\bi^\prime\bSigma^{-1}\bi}\left(\bA_\phi\bi_n-\boldsymbol{\beta}\bi^\prime\bSigma^{-1}\bm\right)\,.
\end{equation}
At last, we put $\bla$ from (\ref{lam}) in (\ref{mvw3}) to get the first statement of Theorem 1.

To show positive definiteness of matrix $\bA$ this we take an arbitrary vector $\bxi\neq\mathbf{0}$ and consider the quadratic form
\begin{equation}
\bxi^\prime\bA_\phi\bxi=\bxi^\prime\bA_0\bB\bxi+\bxi^\prime\bF\bB\bxi+\bar{\phi}(\bxi^\prime\bol{\beta})^2-2\bxi^\prime\bF\bol{\beta}(\bxi^\prime\bol{\beta})\,.
\end{equation}
The first part $\bxi^\prime\bA_0\bB\bxi$ is always positive due to positive definiteness of the matrix $\bA_0\bB$. Now consider the rest denoted by
$Q=\bxi^\prime\bF\bB\bxi+\bar{\phi}(\bxi^\prime\bol{\beta})^2-2\bxi^\prime\bF\bol{\beta}(\bxi^\prime\bol{\beta})$:
\begin{footnotesize}
\begin{eqnarray*}
Q=\sum\limits_{i=1}^n\phi_i\beta_i\xi^2_i+\bar{\phi}\left(\sum\limits_{i=1}^n\beta_i\xi_i\right)^2-2\left(\sum\limits_{i=1}^n\phi_i\beta_i\xi_i\right)\left(\sum\limits_{i=1}^n\beta_i\xi_i\right)
\geq\phi_{min}\left(\sum\limits_{i=1}^n\beta_i\xi^2_i-\left(\sum\limits_{i=1}^n\beta_i\xi_i\right)^2\right)\overset{Jensen}{\geq}0\,.
\end{eqnarray*}
\end{footnotesize}
From the last inequality and $\bxi^\prime\bA_0\bB\bxi>0$ follows $\bxi^\prime\bA_\phi\bxi>0$ and, thus, the positive definiteness of the matrix $1/2(\bA+\bA^\prime)$.

\end{document}